\documentclass[conference]{IEEEtran}
\IEEEoverridecommandlockouts

\usepackage{cite}
\usepackage{amsmath,amssymb,amsfonts}
\usepackage{algorithmic}
\usepackage{algorithm}
\usepackage{graphicx}
\usepackage{caption}
\usepackage{subcaption}
\usepackage{textcomp}
\usepackage{xcolor}
\usepackage{multirow}
\def\BibTeX{{\rm B\kern-.05em{\sc i\kern-.025em b}\kern-.08em
    T\kern-.1667em\lower.7ex\hbox{E}\kern-.125emX}}
\begin{document}

\title{On the Energy Consumption of Rotary Wing and Fixed Wing UAVs in Flying Networks\\
}

\author{\IEEEauthorblockN{Pedro Ribeiro, André Coelho, Rui Campos}
\IEEEauthorblockA{INESC TEC and Faculdade de Engenharia, Universidade do Porto, Portugal\\
\{pedro.m.ribeiro, andre.f.coelho, rui.l.campos\}@inesctec.pt}
}

\maketitle

\begin{abstract}
Unmanned Aerial Vehicles (UAVs) are increasingly used to enable wireless communications. Due to their characteristics, such as the ability to hover and carry cargo, UAVs can serve as communications nodes, including Wi-Fi Access Points and Cellular Base Stations. In previous work, we proposed the Sustainable multi-UAV Performance-aware Placement (SUPPLY) algorithm, which focuses on the energy-efficient placement of multiple UAVs acting as Flying Access Points (FAPs). Additionally, we developed the Multi-UAV Energy Consumption (MUAVE) simulator to evaluate the UAV energy consumption, specifically when using the SUPPLY algorithm. However, MUAVE was initially designed to compute the energy consumption for rotary-wing UAVs only. 

In this paper, we propose eMUAVE, an enhanced version of the MUAVE simulator that allows the evaluation of the energy consumption for both rotary-wing and fixed-wing UAVs. Our energy consumption evaluation using eMUAVE considers reference and random networking scenarios. The results show that fixed-wing UAVs can be employed in the majority of networking scenarios. However, rotary-wing UAVs are typically more energy-efficient than fixed-wing UAVs when following the trajectories defined by SUPPLY.
\end{abstract}


\section{Introduction}
Unmanned Aerial Vehicles (UAVs) have had an increasing interest by the scientific community in the past few years. Due to their capabilities, such as hovering and carrying cargo, UAVs can be used for multiple purposes. A promising application of UAV technology involves its use within the context of wireless communications as part of the so-called Non-Terrestrial Networks (NTNs). NTNs include Flying Networks (FNs) consisting of UAVs carrying communications nodes, such as Wi-Fi Access Points and Cellular Base Stations. FNs can be used to reinforce and assure wireless coverage to Ground Users (GUs), especially in temporary events, such as disaster management scenarios and crowded events in general. Since UAVs mostly rely on onboard power sources that need recharging, there is a need to estimate and optimize the UAV's energy consumption. We previously proposed the Sustainable multi-UAV Performance-aware Placement (SUPPLY) algorithm \cite{Ribeiro2024}. SUPPLY defines energy-efficient trajectories for multiple UAVs acting as Flying Access Points (FAPs). To implement the SUPPLY algorithm and evaluate the energy consumption of an FN when SUPPLY is used, we developed a Python-based simulator named Multi-UAV Energy Consumption (MUAVE) simulator \cite{MUAVE}. MUAVE computes the energy consumption of FAPs for trajectories defined by SUPPLY or any other algorithm implemented in the simulator. Originally, MUAVE was designed to assess only rotary-wing UAVs, limiting its calculations for this type of UAV. This paper expands MUAVE to support fixed-wing UAVs and investigates the energy consumption of fixed-wing UAVs when used as an alternative to rotary-wing UAVs.

The main contributions of this paper are two-fold:

\begin{itemize}
    \item \textbf{The enhanced MUAVE (eMUAVE) simulator}, which is available to the community and supports the computation of the energy consumption for both rotary-wing and fixed-wing UAVs, considering state of the art energy consumption models;
    \item \textbf{The evaluation of energy consumption for fixed-wing UAVs and comparison with rotary-wing UAVs}, when the UAVs follow the trajectories defined by the SUPPLY algorithm.
\end{itemize}

This paper is structured as follows. Section \ref{sec:Background} presents background information on UAV technology and energy consumption models.
Section \ref{sec:SUPPLY} describes the SUPPLY algorithm.
Section \ref{sec:MUAVE} discusses the eMUAVE simulator, including the enhancements introduced. 
Section \ref{sec:Evaluation} presents and discusses the energy consumption results.
Section \ref{sec:Conclusion} summarizes the main conclusions and provides directions for future work.

\section{Background} \label{sec:Background}
\subsection{UAV Types}

\begin{figure*}[h]
\begin{equation}\label{eq:1}
\begin{aligned}
    E(q(t))=& \underbrace{\int_{0}^{T}{P}_{b}\left(1+\frac{3{||v(t)||}^{2}}{{U}_{tip}^2}\right)dt}_{blade\:profile}
    +\underbrace{\int_{0}^{T}P_{i}\sqrt{1+\frac{{a_{c}}^{2}(t)}{g^{2}}}\left( \sqrt{1+\frac{{a_{c}}^{2}(t)}{g^{2}}+\frac{{||v(t)||}^{4}}{4{v}_{0}^{4}}}-\frac{{||v(t)||}^{2}}{2{v}_{0}^{2}} \right)^{1/2}dt}_{induced}\\
    &+\underbrace{\int_{0}^{T}\frac{1}{2}{d}_{0}\rho sA{||v(t)||}^{3}dt}_{parasite}+\Delta_{K}
\end{aligned}
\end{equation}
\end{figure*}

UAVs are the main elements in an FN, carrying communications nodes that provide wireless connectivity to GUs. UAVs can be classified based on their flying mechanism into two types: rotary-wing and fixed-wing.

Rotary-wing UAVs are typically equipped with multiple rotors, each featuring a rotor blade that moves air downwards to generate the necessary lift to keep the UAV airborne. Rotary-wing UAVs are able to hover, making them particularly useful for use cases that require stable wireless coverage. This capability combined with precise trajectory following, allows for optimized positioning of communications nodes in an FN. Additionally, rotary-wing UAVs are versatile in multiple deployment scenarios, as they can perform vertical take-offs and landings without the need for a runway.

Fixed-wing UAVs have rigid wings, similar to airplanes, that generate lift as air passes through them. This flying mechanism necessitates constant UAV movement since it needs to keep moving to generate lift, preventing it from hovering. Fixed-wing UAVs require runways for both take-off and landing, but they excel in flying at greater altitudes and faster speeds while being energy-efficient.

\subsection{UAV Energy Consumption Models}
UAVs, especially within FNs, typically rely on electrical batteries. Due to the limited energy capacity of the batteries, UAVs have limited endurance. For this reason, estimating and optimizing the energy consumption of UAVs is crucial. In FNs, UAVs expend energy for two tasks: communications and propulsion. Since communications account for a small fraction of energy use, they are often excluded from energy consumption models. The scientific community has proposed energy consumption models that realistically replicate the energy and power consumption for both types of UAVs.

Multiple works \cite{Abeywickrama2018, 2Abeywickrama2018, Paredes2017, 8663615, Ding2020, 9461176, 9495369, Ruzicka2023, Muli2022, D'Andrea,  Dorling_2017, Stolaroff2018, Kirschstein2020, Tseng2017} have proposed models for rotary-wing energy consumption. However, this paper will focus on the model already integrated into the MUAVE simulator, which was proposed in \cite{9495369}. This model considers acceleration -- an essential factor for energy consumption -- distinguishing it from other models. It incorporates the centrifugal acceleration, based on the conclusion that tangential acceleration's impact on energy consumption is intrinsically accounted for in the model by changes in velocity.
The model is given by (\ref{eq:1}) and allows computing energy consumption for a UAV's trajectory $q$.

In (\ref{eq:1}), \({P}_{b}\) and \({P}_{ind}\) are two constants representing the \textit{blade profile power} and \textit{induced power}, respectively. \({U}_{tip}\) is the tip speed of the rotor blade, \({v}_{0}\) is the mean rotor induced velocity. \({d}_{0}\) and \(s\) are the fuselage drag ratio and rotor solidity, respectively. \({\rho}\) stands for the air density, and \(A\) denotes the rotor disc area. \({v(t)}\) and \({{a_{c}}^{2}(t)}\) are respectively the flying speed and the centrifugal acceleration at time $t$. For circular movements, this model simplifies into a secondary model, defined in (\ref{eq:2}), where $||v(t)||=V$ and ${a}_{c}=\frac{{V}^{2}}{r}$, with $V$ being the flying speed and $r$ the circular radius.

As outlined in (\ref{eq:2}), for circular movements the power consumption of a UAV depends on the flying speed and the circular radius. 
This model enables the power consumption computation for any combination of flying speed and radius. Furthermore, by utilizing optimization techniques, it is possible to obtain the optimal flying speed that minimizes power consumption for each given specific radius. This detail is especially relevant for the SUPPLY algorithm, which allows minimizing the energy consumption along predefined trajectories.

For fixed-wing UAVs, fewer models have been proposed.
Paredes et al \cite{Paredes2017} developed the model defined in (\ref{eq:3}), which predicts the power consumption of a fixed-wing UAV. This model incorporates several parameters: $W$ is the UAV weight, $S$ is the wing's surface area of the wings, $\rho$ is the air density at a given flying altitude. $C_L$ and $C_D$ are non-dimensional coefficients representing the aerodynamic properties of the wings and propellers of the UAVs.

\begin{equation}\label{eq:2}
\begin{split}
P(V,r) = & \underbrace{{P}_{b}\left(1+\frac{3{V}^{2}}{{U}_{tip}^2}\right)}_{blade\:profile} \\
 & + \underbrace{P_{ind}\sqrt{1+\frac{{V}^{4}}{r^{2}g^{2}}}\left( \sqrt{1+\frac{{V}^{4}}{r^{2}g^{2}}+\frac{{V}^{4}}{4{v}_{0}^{4}}}-\frac{{V}^{2}}{2{v}_{0}^{2}} \right)^{1/2}}_{induced} \\
 & + \underbrace{\frac{1}{2}{d}_{0}\rho sA{V}^{3}}_{parasite}
\end{split}
\end{equation}

\begin{equation}\label{eq:3}
P=\frac{W^{\frac{3}{2}}}{\sqrt{\frac{1}{2}\rho S \frac{C_{L}^{3}}{C_{D}^{2}}}}
\end{equation}

In \cite{Zeng2017}, the authors developed a theoretical model, defined in (\ref{eq:4}), to calculate the propulsion energy consumption of ﬁxed-wing UAVs. The energy consumption over a trajectory $q$ is a function of the speed $v(t)$ and acceleration $a(t)$, with parameters $c_1$ and $c_2$ reflecting the UAV characteristics and environmental conditions such as weight, wing area, and air density.

\begin{figure*}[]
\begin{equation}\label{eq:4}
\begin{aligned}
   E(q(t))=\int_{0}^{T}\left [ c_{1}\left \| v(t) \right \| ^{3}+\frac{c_{2}}{\left \|v(t) \right \|}\left ( 1+\frac{\left \|a(t)  \right \|^{2}-\frac{\left \| \left (a^{T}(t)v(t)  \right ) \right \| )^{2}}{\left \| v(t) \right \|^2}}{g^2} \right) \right ]dt+\frac{1}{2}m\left ( \left \| v(T) \right \| ^{2}-\left \|v(0)  \right \|^2 \right )
\end{aligned}
\end{equation}
\end{figure*}

Similarly to the simplification resulting in the consumption model defined in (\ref{eq:2}), the model in (\ref{eq:4}) simplifies for circular movements. Considering $v(t)=V$, $a^T(t)v(t) = 0$, and ${a}(t)=||a(t)||=\frac{{V}^{2}}{r}$, the derived power consumption model is given by (\ref{eq:5}).

Building upon the model in (\ref{eq:4}), the authors of \cite{Sun2022} and \cite{Xiong2023} proposed an energy consumption model for fixed-wing UAVs following 3D trajectories, considering the 3D velocity and acceleration vector components.

\begin{equation}\label{eq:5}
\begin{aligned}
   P(V,r) = \left ( c_1+\frac{c_2}{g^2r^2} \right )V^3+\frac{c_2}{V}
\end{aligned}
\end{equation}

\section{Sustainable multi-UAV Performance-aware Placement Algorithm}\label{sec:SUPPLY}
In this section, we present the SUPPLY algorithm, initially proposed in \cite{Ribeiro2024}. It is designed for networking scenarios composed of multiple GUs with heterogeneous Quality of Service (QoS) requirements, distributed in a given area. The SUPPLY algorithm groups the GUs to minimize the communications resources (number of FAPs) needed while defining energy-efficient trajectories for FAPs assuring QoS to the GUs.

\begin{figure}[]
 \centering
 \begin{subfigure}[c]{0.24\textwidth}
    \centering
    \includegraphics[width=\textwidth]{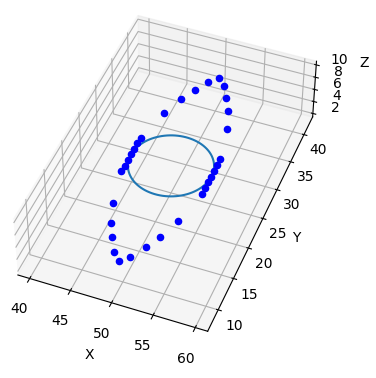}
    \caption{Circular trajectory.}
    \label{fig:Cicular}
 \end{subfigure}
 \hfill
 \begin{subfigure}[c]{0.24\textwidth}
    \centering
    \includegraphics[width=\textwidth]{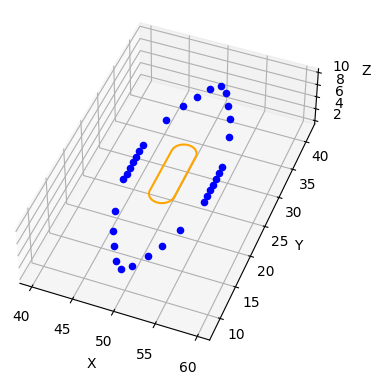}
    \caption{Inner Elliptic trajectory.}
    \label{fig:Ovalcircular}
 \end{subfigure}
\hfill
\begin{subfigure}[c]{0.24\textwidth}
    \centering
    \includegraphics[width=\textwidth]{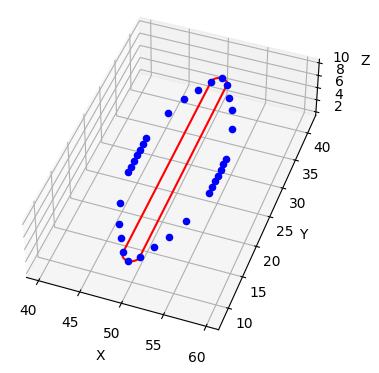}
    \caption{Elliptic trajectory.}
    \label{fig:Ovalarea}
 \end{subfigure}
\caption{Intersection area perimeter and SUPPLY trajectories.}
\label{fig:Trajectories}
\end{figure}

SUPPLY, presented in Algorithm \ref{alg:1}, consists of two phases.
The first phase groups the GUs to minimize the number of FAPs needed. The second phase defines the energy-efficient trajectories for each FAP in the network. The algorithm starts by calculating the Signal-to-Noise Ratio (SNR) for potential links between the GUs and FAPs at all the possible positions. The FAPs are at a fixed altitude of 6 meters to prevent collisions with GUs and optimize SNR. Transmission power is set to 20 dBm, with a 1 dB SNR margin added for reliable network performance.
The SNR values obtained are used to estimate the capacity of the wireless links by correlating them with the tabled data rate associated with each Modulation and Coding Scheme (MCS) index \cite{wireless}. 
Considering the information on the GUs positions, the possible FAP positions, the GUs offered load, and the capacity of the wireless links, the algorithm determines a GUs grouping solution that minimizes the number of FAPs required.
SUPPLY then defines the maximum distance that allows achieving the necessary SNR in the links between the GUs and the FAPs. In 3D space, this corresponds to a sphere around each GU, with the radius set equal to the defined maximum distance.
Each FAP should be positioned in the subspace that results from the intersection of the GUs spheres it is serving. Considering a fixed altitude of 6 meters, each FAP is positioned in an area within this subspace, called the intersection area. 
The algorithm assures that the resulting intersection areas do not overlap, in order to avoid collisions between FAPs, by removing any overlapped areas.
The next step is to determine the centroid and the perimeter of the intersection areas, which are crucial to delineate the trajectories. Subsequently, the SUPPLY algorithm defines three trajectories: Circular, Inner Elliptic, and Elliptic, as depicted in Fig. \ref{fig:Trajectories}. The Circular trajectory consists of a circumference with a radius equal to the minimum distance from the area's centroid to its perimeter. The elliptic trajectories consist of two straight-line segments and two semi-circles. The Inner Elliptic trajectory is confined within the area delimited by the Circular trajectory. The Elliptic trajectory extends to all of the intersection area. 
In the final step, the SUPPLY algorithm selects the most energy-efficient trajectory, which results in the least energy consumption for each FAP. 

\begin{algorithm}
\caption{SUPPLY Algorithm}\label{alg:1}
\hspace*{\algorithmicindent} \textbf{Input}: GUs positions, GUs offered load, possible FAP positions\\
\hspace*{\algorithmicindent} \textbf{Output:} Trajectories of the FAPs\\
\begin{algorithmic}[1]
\FORALL{Possible FAP positions}\label{l:1}
    \STATE Calculate SNR values for potential links with GUs \label{l:2}
    \STATE Estimate capacity of wireless links GU-FAP\label{l:3}
\ENDFOR
\STATE Groups of GUs $\leftarrow$  Optimizer (GUs positions, possible FAP positions, GUs offered load, capacity of the wireless links GU-FAP)\label{l:4}
\FORALL{Groups of GUs}
    \STATE Define minimum target SNR for each GU \label{l:7}
    \STATE Calculate maximum Euclidean distance that ensures target SNR for each GU\label{l:8}
    \STATE Calculate intersection area \label{l:9}
    \IF{Intersection area overlaps with previous calculated intersection's areas} \label{l:10}
        \STATE Remove overlapped area from intersection area to avoid collisions between FAPs \label{l:11}
    \ENDIF
    \STATE Compute centroid of intersection area \label{l:14}
    \STATE Compute perimeter of intersection area \label{l:15}
    \STATE Compute Circular trajectory \label{l:16}
    \STATE Compute Inner Elliptic trajectory \label{l:17}
    \STATE Compute Elliptic trajectory \label{l:18}
    \STATE Choose most energy-efficient FAP trajectory \label{l:19}
\ENDFOR
\end{algorithmic}
\end{algorithm}

\section{Enhanced Multi-UAV Energy Consumption Simulator}\label{sec:MUAVE}
The Enhanced Multi-UAV Energy Consumption (eMUAVE) simulator \cite{eMUAVE} implements the SUPPLY algorithm and computes the energy consumption for multi-UAV networks. eMUAVE, implemented in Python, builds upon MUAVE \cite{MUAVE} and the UAV Power Simulator proposed by Rodrigues H. in \cite{UPS}, which was designed for a single UAV acting as a gateway for a group of FAPs. eMUAVE includes all the original features of MUAVE while augmenting its capabilities to compute energy consumption simulation for fixed-wing UAVs. Moreover, it allows for a comparative analysis of the energy consumption per hour related to the two types of UAVs when following the SUPPLY-defined trajectories for the same scenario; the support for other trajectories can be added to eMUAVE in the future.

eMUAVE considers a set of configuration parameters related to network configurations, UAV specifications, and environmental characteristics. eMUAVE integrates the simplified power consumption model (\ref{eq:2}) for calculating the energy usage of rotary-wing UAVs for a given trajectory, which can include both circular and straight-line flight segments.
For fixed-wing UAVs, the energy consumption model given by (\ref{eq:4}) was selected. This model has been used in multiple state of the art works, such as \cite{Dong2019}, \cite{XIE2022}, and \cite{Hu2022}, and also served as a baseline to develop more complex models for energy consumption in 3D trajectories.
Similar to the model used for rotary-wing UAVs, the simplified version given by (\ref{eq:5}) was considered.
Regarding the integration of this model, additional factors were taken into account due to the flight dynamics and capabilities of fixed-wing UAVs. These UAVs require considering a minimum curve radius and impose a maximum value for the centrifugal acceleration. The original model \cite{Zeng2017} proposes a maximum acceleration of 5 $m/s^2$, a value adopted by later studies, which also considered a minimum radius of 5 $m$.
In its current version, eMUAVE introduces the minimum radius constraint with a value of 5 $m$. While acceleration is not directly limited, as in most state of the art works, maintaining a radius of at least 5 $m$ ensures that the acceleration does not significantly exceed the limits considered in other state of the art works.
A limitation of this model is related to the parameters $c_1$ and $c_2$ used to simulate all the environmental and UAV characteristics. The authors assume $c_1 = 9.26\times10^{-4}$ and $c_2 = 2250$, resulting in a speed $V_{em} = 30 m/s$, which leads to the minimum energy consumption and the corresponding minimum propulsion power consumption $P_{em} = 100 W$ in straight-line flight. Equations to calculate these parameters are provided; the parameters used in those equations are not. Due to the inability to obtain environment and UAV parameters that allow the computation of alternative $c_1$ and $c_2$ values, in eMUAVE the same values were considered. A fair comparison is achieved by configuring the rotary-wing UAVs to have a comparable minimum propulsion power consumption of $P_{em} = 126 W$.

Regarding energy consumption, in each segment of the trajectories with a radius $r$, eMUAVE calculates the optimal UAV flying speed $V$, which leads to the minimum power consumption.
This optimization uses methods from the SciPy package \cite{SciPy}. 
For circular trajectories, a single value for speed and corresponding power consumption is calculated. For elliptic trajectories, which consist of two straight-line segments and two semi-circles, the speed is optimized separately for each segment to ensure the lowest possible power consumption. For the straight-line segments of elliptic trajectories, the model treats radius $r$ as approaching infinity, simplifying the calculation. In order to calculate the energy consumption over a given period, the average power consumption along these trajectories is considered.
For comparison reasons, eMUAVE is also able to calculate the hovering power consumption for rotary-wing UAVs by setting the flying speed $V$ to 0 $m/s$, as well as the optimal steady-state power consumption, maintained by having the FAP moving in a straight line at the optimal constant speed.
By implementing the SUPPLY algorithm, eMUAVE selects the trajectories that minimize the power consumption for each FAP of each UAV type. Finally, it generates a graphical representation of the energy consumption per hour.

As for additional outputs, eMUAVE provides graphical representations of the intersection areas and the three different SUPPLY trajectories, as illustrated in Fig. \ref{fig:Trajectories}. Furthermore, it generates text files documenting the FAPs’ movements over time, along each trajectory. Additionally, eMUAVE outputs generic information, including intermediate results such as calculated radii, power metrics, and the selected trajectories.

\section{UAV Energy Consumption Evaluation}\label{sec:Evaluation}
In order to assess the potential of using fixed-wing UAVs following the SUPPLY trajectories, a simulation-based energy consumption evaluation using the eMUAVE simulator was performed.
The configuration of eMUAVE took into account the UAV, the environment, and the network parameters presented in Table \ref{tab:SimParameters}.

\begin{table}
\caption{Simulation Parameters.}
\label{tab:SimParameters}
\centering
\setlength{\tabcolsep}{3pt}
\begin{tabular}{|p{175 pt}|c|}
\hline
\bfseries Parameter & \bfseries Value  \\ \hline \hline
Rotary-wing UAV weight ($W$) & 20 $N$           \\
Rotor radius ($R$) & 0.4 $m$         \\
Blade angular velocity ($\Omega$) & 300 $rad/s$      \\
Incremental correction factor to induced power ($k$) & 0.1   \\
Profile drag coefficient ($\delta$) & 0.012            \\
Air density ($\rho$) & 1.225 $kg/m^3$     \\
Gravitational acceleration ($g$) & 9.8 $m/s^{2}$    \\
Rotor disc area ($A$) & 0.503 $m^2$        \\ 
Tip speed of the rotor blade ($U_{tip}$) & 120 $m/s$        \\ 
Fuselage drag ratio ($d_{0}$)  & 0.6              \\
Mean rotor induced velocity in hovering state ($v_{0}$) & 4.03             \\ 
Rotor solidity ($s$) & 0.05             \\ 
Blade profile power in hovering state ($P_{b}$) & 79.86            \\ 
Induced power in hovering state ($P_{ind}$) & 88.63 \\
Fixed-wing parameter 1 ($c_1$) & $9.26\times10^{-4}$ \\ 
Fixed-wing parameter 2 ($c_2$) & 2250 \\
Fixed-wing minimum radius & 5 $m$ \\
Wi-Fi Standard& IEEE 802.11ac        \\ 
Channel Bandwidth & 160 $MHz$       \\ 
Wireless channel & 50        \\
Channel frequency & 5250 $MHz$      \\
Guard interval & 800 $ns$         \\
Transmission power & 20 $dBm$ \\
Noise power & -85 $dBm$ \\
\hline
\end{tabular}
\end{table}

\subsection{Energy Evaluation Under Reference Networking Scenarios}
Multiple scenarios were analyzed to evaluate the effectiveness of fixed-wing UAVs in following the defined trajectories and compare the energy consumption between the two UAV types. The scenarios consist of a variable number of GUs with heterogeneous QoS requirements, randomly distributed in an area of 100 × 100 meters. For each number of GUs in the area (2, 5, and 10 GUs), one scenario was analyzed. The scenarios were selected so that the energy consumption comparison between the two UAV types was possible. The evaluated scenarios are characterized in Table \ref{tab:NetworkScenarios}.

\renewcommand{\arraystretch}{1.5}
\begin{table*}[]
\caption{GUs positions and GUs offered load for the networking scenarios considered.}
\label{tab:NetworkScenarios}
\centering
\begin{tabular}{|c|c|c|c|} %p{0.2\textwidth}
\hline
\multicolumn{1}{|c|}{\textbf{Network Scenario}} & \textbf{GUs Positions (x,y,z)}  & \textbf{GUs Offered Load (Mbit/s)}   \\  \hline \hline
2 GUs  & (47,32,0), (52,71,0) &  200, 117             \\ \hline 
\multirow{2}{*}{5 GUs} \multirow{2}{*}      & (19,62,0), (85,46,0), & 36, 27,                     \\ 
                                            & (86,53,0), (2,9,0), (52,88,0) & 19, 14, 23          \\ \hline  
\multirow{3}{*}{10 GUs} \multirow{3}{*}       & (69,83,0), (68,91,0), (26,16,0), & 9, 6, 1,            \\ 
                                              & (67,8,0), (38,21,0), (23,71,0), & 5, 3, 6,             \\
                                              & (60,34,0), (8,31,0), (59,59,0), (20,79,0) &  7, 5,
8, 6\\ 
\hline
\end{tabular}
\end{table*}

\begin{figure}
    \centering
    \includegraphics[width=0.49\textwidth]{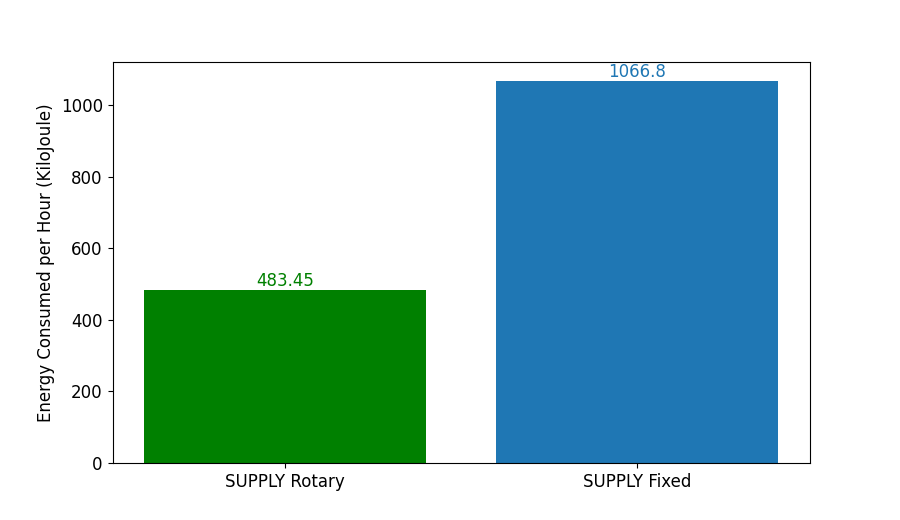}
    \caption{Comparison of energy consumption per hour for a reference scenario with 2 GUs}
    \label{fig:2G1F}
\end{figure}

The first scenario considered 2 GUs. The comparison results for the energy consumption per hour are presented in Fig. \ref{fig:2G1F}. SUPPLY selected the circular trajectory for both UAV types. The energy consumption per hour for rotary-wing and fixed-wing UAVs was 483 kJ and 1067 kJ, respectively, showing an increase of more than 100\% when using a fixed-wing UAV.

\begin{figure}
    \centering
    \includegraphics[width=0.49\textwidth]{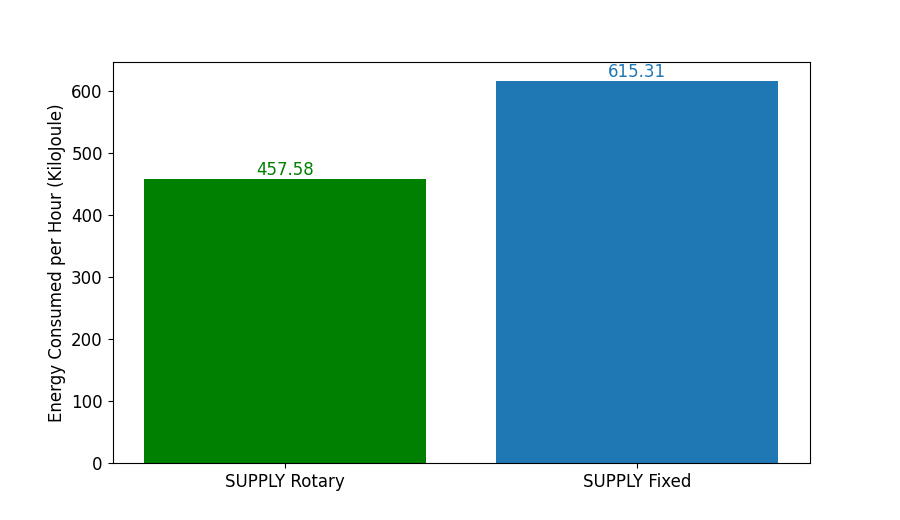}
    \caption{Comparison of energy consumption per hour for a reference scenario with 5 GUs.}
    \label{fig:5G1F}
\end{figure}

The second scenario considered 5 GUs. SUPPLY again chose the circular trajectory for both UAV types. The comparison results for the energy consumption per hour are presented in Fig. \ref{fig:5G1F}. The energy consumption per hour results are equal to 458 kJ for rotary-wing UAVs and 615 kJ for fixed-wing. Despite a considerably smaller difference, this still represents an increase of 34\% in energy consumption for the fixed-wing UAV. 

\begin{figure}
    \centering
    \includegraphics[width=0.49\textwidth]{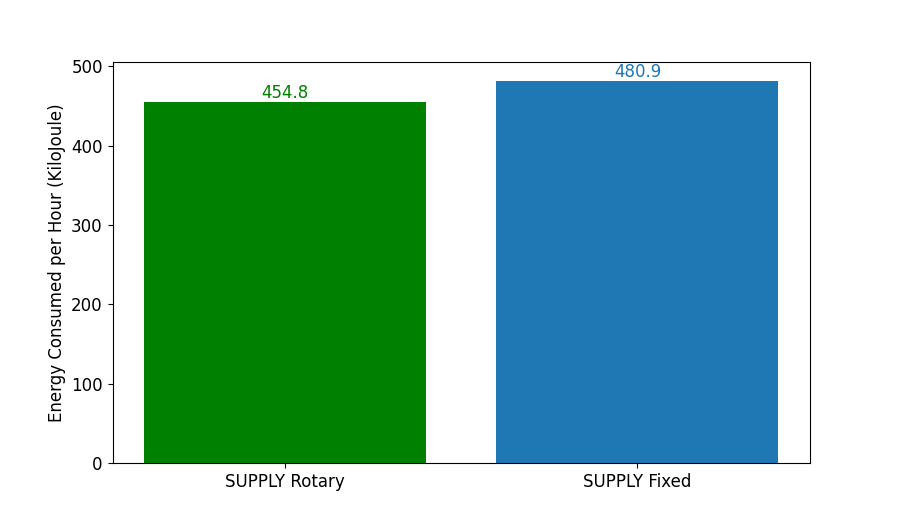}
    \caption{Comparison of energy consumption per hour for a reference scenario with 10 GUs.}
    \label{fig:10G1F}
\end{figure}

The third scenario considered 10 GUs. SUPPLY selected the circular trajectory for the two UAV types. Fig. \ref{fig:10G1F} presents the comparison results for the energy consumption per hour. As shown, similar results were obtained for rotary-wing and fixed-wing UAVs: 455 kJ and 481 kJ, respectively. Due to the large radius of the circular trajectory ($r$ = 108 m), the power consumption along the trajectory was almost identical. 

\subsection{Energy Evaluation for Random Networking Scenarios}
\begin{figure}
    \centering
    \includegraphics[width=0.49\textwidth]{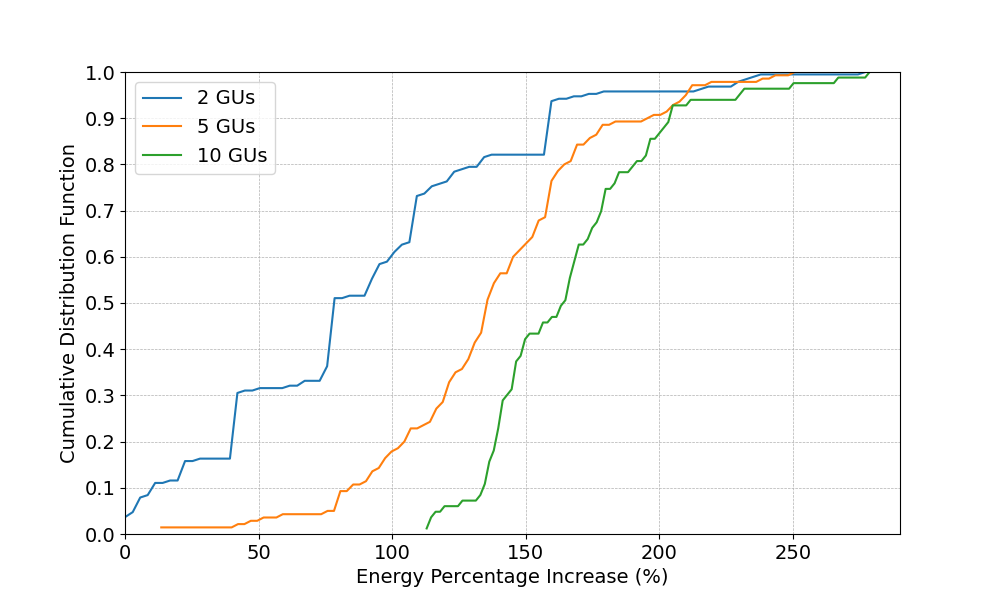}
    \caption{Percentage increase in energy consumption per hour of fixed-wing UAVs compared to rotary-wing UAVs, for each distinct number of GUs – 2, 5, and 10.}
    \label{fig:200}
\end{figure}

Considering randomly generated networking scenarios with a predefined number of GUs and a predefined interval for the GUs' offered loads, we compared the energy consumption per hour of fixed-wing UAVs relative to rotary-wing UAVs. We also analyzed how the number of GUs in the network affects the energy consumption. From the evaluation for reference networking scenarios, we observed that using fixed-wing UAVs tended to lead to higher energy consumption and that fixed-wing UAVs were not always possible to employ. As such, we focused on understanding: 1) the increase in energy consumption when compared to rotary-wing UAVs and 2) the percentage of scenarios where the use of fixed-wing UAVs was impossible. For this purpose, we generated 200 random scenarios for different number of GUs in the network: 2, 5, and 10 GUs. The GUs' offered load values were randomly selected within an interval from 0 to 500 Mbit/s for all scenarios. Fig. \ref{fig:200} presents the percentage increase in energy consumption per hour compared to rotary-wing UAVs for each number of GUs in the network. The $50^{th}$ percentile reveals an energy percentage increase of $75\%$, $134\%$, and $163\%$ for scenarios with 2, 5, and 10 GUs, respectively. It is important to note that the results presented in Fig. \ref{fig:200} comprise only the networking scenarios where it was possible to employ both UAV types. The scenarios with 2, 5, and 10 GUs showed that using fixed-wing UAVs was impossible in 5\%, 30\%, and 59\% of cases, respectively.

\subsection{Discussion}
From the results for reference networking scenarios, we conclude that the SUPPLY algorithm achieves better energy consumption results with rotary-wing UAVs in all evaluated scenarios. Although fixed-wing UAVs can be used in many scenarios, their utilization leads to significantly worse energy consumption results. Furthermore, due to the limitation introduced by their flying mechanism, fixed-wing UAVs can not always follow the trajectories defined by the SUPPLY algorithm. This becomes apparent when smaller intersection areas are defined. Additionally, the elliptic SUPPLY trajectories (Elliptic and Inner Elliptic) are not suitable for fixed-wing UAVs, as they result either in higher power consumption or are impossible to employ due to the radius constraint.
When considering smaller radii, the power consumption of fixed-wing UAVs increases significantly more than that of rotary-wing UAVs. However, it is worth noting that for large enough intersection areas that lead to large radii ($r$ $>$ 100 m), fixed-wing UAVs can achieve similar energy consumption results compared to rotary-wing UAVs. 

From the evaluation for random networking scenarios, we verify that the SUPPLY algorithm generally achieves better energy consumption results with rotary-wing UAVs. Furthermore, increasing the number of GUs leads to a general increase in energy consumption for both UAV types. As the number of GUs in the same area increases, the SUPPLY algorithm tends to generate larger groups, which result in smaller intersection areas. With a smaller area to place the FAP, the SUPPLY algorithm defines less energy-efficient trajectories, characterized by smaller straight-line segments and radii.
Fixed-wing UAVs experience a higher energy consumption increase as the reduction in trajectory curve radius has a higher negative impact for this UAV type. Additionally, there is a higher number of impossible scenarios for fixed-wing UAVs with the increase in the number of GUs, as more trajectories fail to meet the radius constraint imposed by this UAV type.

\section{Conclusion}\label{sec:Conclusion}

In this paper, we presented eMUAVE, a simulator capable of computing the energy consumption of rotary-wing and fixed-wing UAVs, considering state of the art energy consumption models. eMUAVE allowed us to evaluate and compare the energy consumption of both UAV types when following trajectories defined by the SUPPLY algorithm.

Regarding the suitability of fixed-wing UAVs, although they can be used in most scenarios, the energy consumption results are typically worse than those observed with rotary-wing UAVs. Moreover, due to the constraints of their flying mechanism, fixed-wing UAVs are not capable of hovering and often cannot follow the SUPPLY trajectories. For these reasons, rotary-wing UAVs are more suitable to operate in FNs that use the SUPPLY algorithm.

For future research, exploring other networking scenarios may reveal conditions where the integration of fixed-wing UAVs is more favorable. These scenarios may include data distribution or other scenarios where UAVs are not required to provide continuous service to GUs and thus do not have to follow trajectories with extreme curve radii.

\bibliographystyle{IEEEtran}
\bibliography{refs}

\end{document}